%% file: EUSIPCO_paper_IEEE_template.tex
\documentclass[conference,10pt,a4paper,twocolumn,oneside]{IEEEtran_doc_class}
\IEEEoverridecommandlockouts
\usepackage{graphicx}
\usepackage[usenames,dvipsnames]{xcolor}
\usepackage{pgfplots}
\usepackage{amssymb}
\usepackage{amsmath}
\usepackage{bm}
\usepackage{amsfonts}
\usepackage{epstopdf}
\usepackage{subfig}
\usepackage{cite}
\usepackage{tikz}
\usepackage{booktabs}
\usepackage{algorithmic}
\usepackage{algorithm}
\usetikzlibrary{plotmarks}
\begin{document}
\bstctlcite{IEEEexample:BSTcontrol}
%
\title{A Variational EM Method for Pole-Zero \\ Modeling of Speech with Mixed Block Sparse and Gaussian Excitation}

\author{\IEEEauthorblockN{Liming Shi\thanks{This work was funded by the Danish Council for Independent Research, grant ID: DFF 4184-00056}, Jesper Kj\ae r Nielsen, Jesper Rindom Jensen and Mads Gr\ae sb\o ll Christensen}
\IEEEauthorblockA{Audio Analysis Lab, AD:MT, Aalborg University\\
Emails: \{ls, jkn, jrj, mgc\}@create.aau.dk}}


\maketitle

\begin{abstract}
The modeling of speech can be used for speech synthesis and speech recognition. We present a speech analysis method based on pole-zero modeling of speech with mixed block sparse and Gaussian excitation. By using a pole-zero model, instead of the all-pole model, a better spectral fitting can be expected. Moreover, motivated by the block sparse glottal flow excitation during voiced speech and the white noise excitation for unvoiced speech, we model the excitation sequence as a combination of block sparse signals and white noise. A variational EM (VEM) method is proposed for estimating the posterior PDFs of the block sparse residuals and point estimates of modelling parameters within a sparse Bayesian learning framework. Compared to conventional pole-zero and all-pole based methods, experimental results show that the proposed method has lower spectral distortion and good performance in reconstructing of the block sparse excitation.
\end{abstract}

\section{Introduction}
The modeling of speech has important applications in speech analysis \cite{Makhoul1975}, speaker verification \cite{Pohjalainen2014}, speech synthesis\cite{Erro2014}, etc. Based on the source-filter model, speech is modelled as being produced by a pulse train or white noise for voiced or unvoiced speech, which is further filtered by the speech production filter (SPF) that consists of the vocal tract and lip radiation.

All-pole modeling with a least squares cost function performs well for white noise and low pitch excitation. However, for high pitch excitation, it leads to an all-pole filter with poles close to the unit circle, and the estimated spectrum has a sharper contour than desired \cite{Murthi2000, Drugman2014}. To obtain a robust linear prediction (LP), the Itakura-Saito error criterion \cite{El-Jaroudi1991}, the all-pole modeling with a distortionless response at frequencies of harmonics \cite{Murthi2000}, the regularized LP \cite{Ekman2008} and the short-time energy weighted LP \cite{Alku2013} were proposed. Motivated by the compressive sensing research, a least 1-norm criterion is proposed for voiced speech analysis \cite{Giacobello2012}, where sparse priors on both the excitation signals and prediction coefficients are utilized.  Fast methods and the stability of the 1-norm cost function for spectral envelope estimation are further investigated in \cite{Giacobello2014, Jensen2016}. More recently, in \cite{Giri2014}, the excitation signal of speech is formulated as a combination of block sparse and white noise components to capture the block sparse or white noise excitation separately or simultaneously. An expectation-maximization (EM) algorithm is used to reconstruct the block sparse excitation within a sparse Bayesian learning (SBL) framework \cite{Tipping2001}. 
 
A problem with the all-pole model is that some sounds containing spectral zeros with voiced excitation, such as nasals, or laterals, are poorly estimated by an all-pole model but trivial with a pole-zero (PZ) model \cite{Stoica2004a, Marelli2010}. The estimation of the coefficients of the pole-zero model can be obtained separately \cite{Durbin1960}, jointly \cite{Levy1959} or iteratively \cite{Steiglitz1977}. A 2-norm minimization criterion with Gaussian residuals assumption is commonly used. Frequency domain fitting methods based on a similarity measure is also proposed. Motivated by the logarithmic scale perception of the human auditory system, the logarithmic magnitude function minimization criterion has been proposed \cite{Kobayashi1990, Marelli2010}. In \cite{Kobayashi1990}, the nonlinear logarithm cost function is solved by transforming it into a weighted least squares problem. The Gauss-Newton and Quasi-Newton methods for solving it are further investigated in \cite{Marelli2010}. To consider both the voiced excitation and the PZ model, a speech analysis method based on the PZ model with sparse excitation in noisy conditions is presented \cite{shi2017}. A least 1-norm criterion is used for the coefficient estimation, and sparse deconvolution is applied for deriving sparse residuals.
 
In this paper, we propose a speech analysis method based on the PZ model with mixed excitation. Using the mixed excitation and PZ modeling together, we combine the advantages of non-sparse and sparse algorithms, and obtain a better fitting for both the excitation and SPF spectrum. Using the PZ model, instead of the all-pole model, a better spectral fitting can be expected. Moreover, we model both the voiced, the unvoiced excitation or a mixture of them by the mixed excitation. Additionally, block sparsity is imposed on the voiced excitation component, motivated by the quasi-periodic and temporal-correlated nature of the glottal excitation \cite{Alku2011, Giri2014}. The posterior probability density functions (PDFs) for the sparse excitation and hyper-parameters, as well as point estimates of the PZ model parameters are obtained using the VEM method. 
\section{Signal models}
\label{sec:Signal models}
Consider the following general speech observation model:
\begin{equation} \label{eq1}
y\left ( n \right ) = s\left ( n \right )  + u(n),
\end{equation}
where $y(n)$ is the observation signal and $u(n)$ denotes the noise. We assume that the clean speech signal $s(n)$ is produced by the PZ speech production model, i.e.,  
\begin{equation} \label{eq2}
s\left ( n \right ) = -\sum_{k= 1}^{K}a_k s\left ( n-k \right ) + \sum_{l= 0}^{L}b_le\left ( n-l \right )+ m(n), 
\end{equation}
where $a_k$ and $b_l$ are the modeling coefficients of the PZ model with $b_0=1$, $e\left ( n \right )$ is a sparse excitation corresponding to the voiced part and $m(n)$ is the non-sparse Gaussian excitation component corresponding to the unvoiced part. Assuming $s\left ( n \right )=0 \ \mathrm{for} \  n \leq 0$ and considering one frame of speech signals of $N$ samples, \eqref{eq1} and \eqref{eq2} can be written in matrix forms as 
\begin{align}
\mathbf{y}=&\mathbf{s}+\mathbf{u},  \label{eq3} \\ 
\mathbf{A}\mathbf{s}=&\mathbf{B}\mathbf{e}+\mathbf{m}, \label{eq4} 
\end{align}
where $\mathbf{A}$ and $\mathbf{B}$ are the $N \times {N}$ lower triangular Toeplitz matrices with $[1, a_1, a_2, \cdots, a_{K}, 0, \cdots, 0]$ and $[1, b_1, b_2, \cdots, b_{L}, 0, \cdots, 0]$ as the first columns, respectively.
The block sparse residuals are defined as $\mathbf{e}=[e\left ( 1 \right ) \cdots e\left (  {N} \right )]^{T}$,  and $\mathbf{m}$, $\mathbf{s}$, $\mathbf{y}$ and $\mathbf{u}$ are defined similarly to $\mathbf{e}$. When $L=0$, $\mathbf{B}$ reduces to the identity matrix and \eqref{eq4} becomes the all-pole model. Combining \eqref{eq3} and \eqref{eq4}, the noisy observation can be written as
\begin{equation}\label{eq5}
\mathbf{A}\mathbf{y}=\mathbf{B}\mathbf{e}+\mathbf{m}+\mathbf{A}\mathbf{u}.
\end{equation}
In \cite{shi2017}, we assumed that only the sparse excitation was present ($\mathbf{m}=\mathbf{0}$, but $\mathbf{u} \neq \mathbf{0}$). The sparse residuals and model parameters were estimated iteratively. The sparse residuals were obtained by solving
\begin{equation} \label{eq6}
\min\limits_{\mathbf{e}} \left \| \mathbf{e} \right \|_{1}^{1}   \,\,\,  \text{s.t.}  \,\,\,
\left \| \mathbf{y}-\mathbf{A}^{-1}\mathbf{B}\mathbf{e} \right \|_{2}^{2} \leq C.
\end{equation} 
where $C$ is a constant proportional to the variance of the noise. The model parameters was estimated using the $l_{1}$ norm of the residuals as the cost function (see \cite{shi2017} for details).
\section{Proposed Variational EM method} 
We now proceed to consider the noise-free scenario but with mixed excitation ($\mathbf{u}= \mathbf{0}$, but $\mathbf{m}\neq \mathbf{0}$ ). We consider the pole-zero model parameters $\mathbf{a}=  [a_1,a_2, \cdots, a_K]^{T}$ and $\mathbf{b}=  [b_1, b_2, \cdots, b_L]^{T}$ to be deterministic but unknown. Utilizing the SBL \cite{Tipping2001} methodology, we first express the hierarchical form of the model as
\begin{align}  \label{eq7}
& \;\;\;\;\;\; \mathbf{A}\mathbf{y}=  \mathbf{B}\mathbf{e}+\mathbf{m},  \;\;\;\;\;\;\;\;\; \mathbf{m}  \sim  \mathcal{N}(\mathbf{0}, \gamma_m^{-1}\mathbf{I}_N), \nonumber \\
&\mathbf{e}  \sim  \mathcal{N}(\mathbf{0}, \bm{\Gamma}_{e}^{-1}),  \;\;\; \gamma_m  \sim  \Gamma(c, d), \;\;\; \bm{\alpha} \sim  \prod_{o=1}^{ {O}} \Gamma(\alpha_o; e, f) ,
\end{align}
where ${O}$ is the number of blocks, $\mathbf{\Gamma}_{e}= \text{diag}(\bm{\alpha})\otimes \mathbf{I}_{ {D}}$,  $\otimes$ is the Kronecker product, ${D}$ is the block size, $ {N}= {D} {O}$, $\mathcal{N}$ denotes the multivariate normal PDF and ${\Gamma}$ is the Gamma PDF. The hyperparameter $\alpha_{o}$ is the precision of the $o^{\mathrm{th}}$ block, and when it is infinite, the $o^{\mathrm{th}}$ block will be zero. Note that it is trivial to extend the proposed method to any ${D}$. Moreover, when ${D}=1$, each element in $\mathbf{e}$ is inferred independently. Here, block sparsity model is used to take the quasi-periodic and temporal-correlated nature of the voiced excitation into account. The $\mathbf{m}$ is used for capturing the white noise excitation from unvoiced speech frame or a mixture of phonations.

Our objective is to obtain the posterior densities of $\mathbf{e}$, $\gamma_m$ and $\bm{\alpha}$, and point estimates of the model parameters in $\mathbf{a}$ and $\mathbf{b}$. First, we write the complete likelihood, i.e.,
\begin{align} \label{eq8}
p(\mathbf{y}, \mathbf{e}, \bm{\alpha}, \gamma_m ) = & \; p(\mathbf{y}|\mathbf{e}, \gamma_m) p(\mathbf{e}|\bm{\alpha})p(\bm{\alpha})p(\gamma_m) \nonumber \\
= & \; \mathcal{N}(\mathbf{A}\mathbf{y}| \mathbf{B}\mathbf{e}, \gamma_m^{-1}\mathbf{I}_N) \mathcal{N}(\mathbf{e}|0, \mathbf{\Gamma}_{e}^{-1}) \nonumber \\
&  \times \; \Gamma(\gamma_m ; c, d) \prod_{o=1}^{ {O}} \Gamma(\alpha_o; e, f),
\end{align}
where we used $\mathcal{N}(\mathbf{y}| \mathbf{A}^{-1}\mathbf{B}\mathbf{e}, \gamma_m^{-1}(\mathbf{A}^{T}\mathbf{A})^{-1})=\mathcal{N}(\mathbf{A}\mathbf{y}| \mathbf{B}\mathbf{e}, \gamma_m^{-1}\mathbf{I}_N)$ when $ \mathrm{det}(\mathbf{A})=1$. Instead of finding the joint posterior density $p( \mathbf{e}, \bm{\alpha}, \gamma_m |\mathbf{y})$, which is intractable, we adopt the variational approximation \cite{Bishiop2006}. Assume that $p( \mathbf{e}, \bm{\alpha}, \gamma_m |\mathbf{y})$ is approximated by the density $q( \mathbf{e}, \bm{\alpha}, \gamma_m)$, which may be fully factorized as 
\begin{align} \label{eq9}
q( \mathbf{e}, \bm{\alpha}, \gamma_m) =\; q( \mathbf{e})q( \gamma_m)\prod_{o=1}^{O}q( {\alpha_0}),
\end{align}
where the factors are found using an EM-like algorithm \cite{Bishiop2006}.

In the E-step of the VEM method, we fix the model parameters $\mathbf{a}$ and $\mathbf{b}$, and re-formulate the posterior PDFs estimation problem as maximizing the variational lower bound
\begin{align} \label{eq10}
\max\limits_{q } \mathbb{E}_{q}[\log p(\mathbf{y}, \mathbf{e}, \bm{\alpha}, \gamma_m )]+H[q],
\end{align}
where $q$ is the shorthand for $q( \mathbf{e}, \bm{\alpha}, \gamma_m)$, $H[q] $ is defined as $H[q] = - \mathbb{E}_{q}[\log (q)]$, and $\mathbb{E}_{q(x)}[f(x)]$ denotes the expectation of $f(x)$ w.r.t.\ the random variable $x$ (i.e., $\mathbb{E}_{q(x)}[f(x)]=\int f(x) q(x) dx$). Substituting \eqref{eq8} and \eqref{eq9} into \eqref{eq10}, and following the derivation from \cite{Bishiop2006}, we obtain
\begin{align} \label{eq11}
q( \mathbf{e}) &\propto  e^{\mathbb{E}_{q(\bm{\alpha}, \gamma_m)}[\log \mathcal{N}(\mathbf{A}\mathbf{y}| \mathbf{B}\mathbf{e}, \gamma_m^{-1}\mathbf{I}_N) \mathcal{N}(\mathbf{e}|0, \text{diag}(\bm{\alpha})^{-1}\otimes \mathbf{I}_{ {D}})]}, \nonumber \\
q(\alpha_o) &\propto  \Gamma(\alpha_o; e, f)e^{\mathbb{E}_{q(\mathbf{e})}[\log \mathcal{N}(\mathbf{e}|0, \text{diag}(\bm{\alpha})^{-1}\otimes \mathbf{I}_{ {D}})]}, 1 \leq o \leq  {O}, \nonumber \\
q(\gamma_m) &\propto   \Gamma(\gamma_m; c, d)e^{\mathbb{E}_{q(\mathbf{e})}[\log \mathcal{N}(\mathbf{A}\mathbf{y}| \mathbf{B}\mathbf{e}, \gamma_m^{-1}\mathbf{I}_N)]}. 
 \end{align}
It is clearly seen that $q( \mathbf{e})$ in \eqref{eq11} is a Gaussian PDF, i.e.,
\begin{align} \label{eqnew}
q( \mathbf{e})& = \mathcal{N}(\bm{\tilde{\mu}}, \bm{\tilde{\Sigma}}),
\end{align}
where $\bm{\tilde{\Sigma}}=(\mathbb{E}[\gamma_m]\mathbf{B}^{T}\mathbf{B}+\mathbb{E}[\mathbf{\Gamma}_e])^{-1}$ and $\bm{\tilde{\mu}}=\mathbb{E}[\gamma_m]\bm{\tilde{\Sigma}}\mathbf{B}^{T}\mathbf{A}\mathbf{y}$. We also define the auto-correlation matrix $\mathbf{\tilde{R}}=\bm{\tilde{\Sigma}}+\bm{\tilde{\mu}}\bm{\tilde{\mu}}^{T}$. The posterior PDF of $\alpha_o$ in \eqref{eq11} is a Gamma probability density, i.e.,
\begin{align} \label{eqnew2}
q(\alpha_o) = \Gamma( \tilde{e}_o, \tilde{f}_o),
\end{align}
where $\tilde{e}_o=e+D/2$, $\tilde{f}_o=f+\sum_{i=(o-1)D+1}^{oD}\mathbf{\tilde{R}}_{i, i}/2$ and $\mathbf{\tilde{R}}_{i, i}$ denotes the $(i, i)$ element of $\mathbf{\tilde{R}}$. The expectation of the precision matrix is $\mathbb{E}[\mathbf{\Gamma}_e]= \mathrm{diag}(\tilde{e}_1/\tilde{f}_1, \cdots, \tilde{e}_ {O}/\tilde{f}_ {O})\otimes \mathbf{I}_{ {D}}$.
Similar to $\alpha_o$, the posterior PDF of $\gamma_m$ is
\begin{align} \label{eqnew3}
q(\gamma_m) = \Gamma( \tilde{c}, \tilde{d}),
\end{align}
where $\tilde{c}=c+ {N}/2$, $\tilde{d}=d+( \mathrm{tr}(\bm{\tilde{\Sigma}}\mathbf{B}^{T}\mathbf{B})+\|\mathbf{A}\mathbf{y}-\mathbf{B}\bm{\tilde{\mu}}\|_2^{2})/2$. The expectation of $\gamma_m$ can be expressed as $\mathbb{E}[\gamma_m]= \tilde{c}/\tilde{d}$.

In the M-step, we maximize the lower bound \eqref{eq10} w.r.t. the modeling parameters $\mathbf{a}$ and $\mathbf{b}$, respectively. The optimization problems can be shown to be equivalent to $\min\limits_{\mathbf{a} }\mathbb{E}_{q(\mathbf{e})}\|\mathbf{A}\mathbf{y}-\mathbf{B}\mathbf{e}\|_2^{2} \ \mathrm{and} \
\min\limits_{\mathbf{b} }\mathbb{E}_{q(\mathbf{e})}\|\mathbf{A}\mathbf{y}-\mathbf{B}\mathbf{e}\|_2^{2}
$, respectively. To obtain the estimate for $\mathbf{a}$, we first note that $\mathbf{Ay}$ can be expressed as $\mathbf{Ay}=\mathbf{C}\mathbf{a}+\mathbf{y}$, where $\mathbf{C}$ is a $ {N} \times {K}$ Toeplitz matrix of the form
$$\mathbf{C}= 
\left[               
\begin{array}{ccccccc}   
0  &\cdots & 0\\  
y(1) & \ddots & \vdots \\ 
\vdots & \ddots & 0\\
\vdots &  & y(1)\\
\vdots &  & \vdots\\
y( {N}-1)  & \cdots & y( {N}-K)  \\ 
\end{array}
\right]_{ {N} \times {K}}
$$
Using this expression and $q(\mathbf{e})$ obtained in the E-step, the minimization problem w.r.t.\ $\mathbf{a}$ can be re-formulated as 
\begin{align} \label{eq12}
\min\limits_{\mathbf{a} }\mathbb{E}_{q(\mathbf{e})}\|\mathbf{A}\mathbf{y}-\mathbf{B}\mathbf{e}\|_2^{2} \!\!\! \iff &  \!\!\! 
\min\limits_{\mathbf{a} }\|\mathbf{A}\mathbf{y}-\mathbf{B}\bm{\tilde{\mu}}\|_2^{2}+  {tr}(\bm{\tilde{\Sigma}}
\mathbf{B}^{T}\mathbf{B})\nonumber  \\
\iff & \!\!\! \min\limits_{\mathbf{a} }\|(\mathbf{B}\bm{\tilde{\mu}}-\mathbf{y})-\mathbf{Ca}\|_2^{2}.
\end{align} 
As can be seen, \eqref{eq12} is the standard least squares problem and has the analytical solution as
\begin{align} \label{eq13}
\mathbf{a}=(\mathbf{C}^{T}\mathbf{C})^{-1}\mathbf{C}^{T}(\mathbf{B}\bm{\tilde{\mu}}-\mathbf{y}).
\end{align}
We can obtain the solution of $\mathbf{b}$, like $\mathbf{a}$, by setting the derivative of ${\mathbb{E}_{q(\mathbf{e})}\|\mathbf{A}\mathbf{y}-\mathbf{B}\mathbf{e}\|_2^{2}}$ w.r.t.\ $\mathbf{b}$ to zero, i.e., 
\begin{align} \label{eq14}
\frac{\partial \mathbb{E}_{q(\mathbf{e})} \|\mathbf{A}\mathbf{y}-\mathbf{B}\mathbf{e}\|_2^{2}}{\partial \mathbf{b}}
= & \mathbb{E}_{q(\mathbf{e})}[2\mathbf{F}^{T}(\mathbf{Ay}-\mathbf{Be})] \nonumber \\
= & \mathbf{0}_{L \times 1}.
\end{align}
where $\mathbf{F}$ is an $ {N} \times {L}$ lower triangular Toeplitz matrix of the form
$$\mathbf{F}= 
\left[               
\begin{array}{ccccccc}   
0  &\cdots & 0\\  
e(1) & \ddots & \vdots \\ 
\vdots & \ddots & 0\\
\vdots &  & e(1)\\
\vdots &  & \vdots\\
e( {N}-1)  & \cdots & e( {N}-L)  \\ 
\end{array}
\right]_{ {N} \times {L}}
$$
From \eqref{eq14}, we obtain the estimate of $\mathbf{b}$, i.e., 
\begin{align} \label{eq15}
\mathbf{b}=(\mathbb{E}_{q(\mathbf{e})}[\mathbf{F}^{T}\mathbf{F}])^{-1}(\mathbb{E}_{q(\mathbf{e})}[\mathbf{F}^{T}]\mathbf{A}\mathbf{y}-\mathbb{E}_{q(\mathbf{e})}[\mathbf{F}^{T}\mathbf{e}]),
\end{align}
where $\mathbb{E}_{q(\mathbf{e})}[\mathbf{F}^{T}\mathbf{F}]$ is an $L \times L$ symmetric matrix with the $(i, j)^{\mathrm{th}}, j\geq i$ element given by $\sum_{k=1}^{ {N}-j}\mathbf{\tilde{R}}_{k, k+j-i}$. The $\mathbb{E}_{q(\mathbf{e})}[\mathbf{F}^{T}]$ can be obtained by simply replacing the stochastic variable $e(n), 1\leq n \leq  {N}-1$ in $\mathbf{F}^{T}$ with the mean estimate $\tilde{\mu}(n)$ (the $n^{\mathrm{th}}$ element in $\bm{\tilde{\mu}}$). The $\mathbb{E}_{q(\mathbf{e})}[\mathbf{F}^{T}\mathbf{e}]$ is an $L \times 1$ vector with the $l^{\mathrm{th}}$ element given by $\sum_{k=1}^{ {N}-l}\mathbf{\tilde{R}}_{k, k+l}$. Note that the estimation of $\mathbf{b}$ in \eqref{eq15} requires the knowledge of $\mathbf{a}$ and vice versa (see \eqref{eq13}). This coupling is solved by replacing them with their estimates from previous iteration. The algorithm is initialized with $\mathbf{a}=  [1,0, \cdots, 0_K]^{T}$, $\mathbf{b}=  [1, 0, \cdots, 0_L]^{T}$, $\gamma_m=10$ and $\alpha_{o}=1, o=1,\cdots, O$, and starts with the update of $\mathbf{e}$. We refer to the proposed variational expectation maximization pole-zero estimation algorithm as the VEM-PZ.

\section{Results}
\label{sec:results}

\begin{figure}[tb]
\centering
\includegraphics{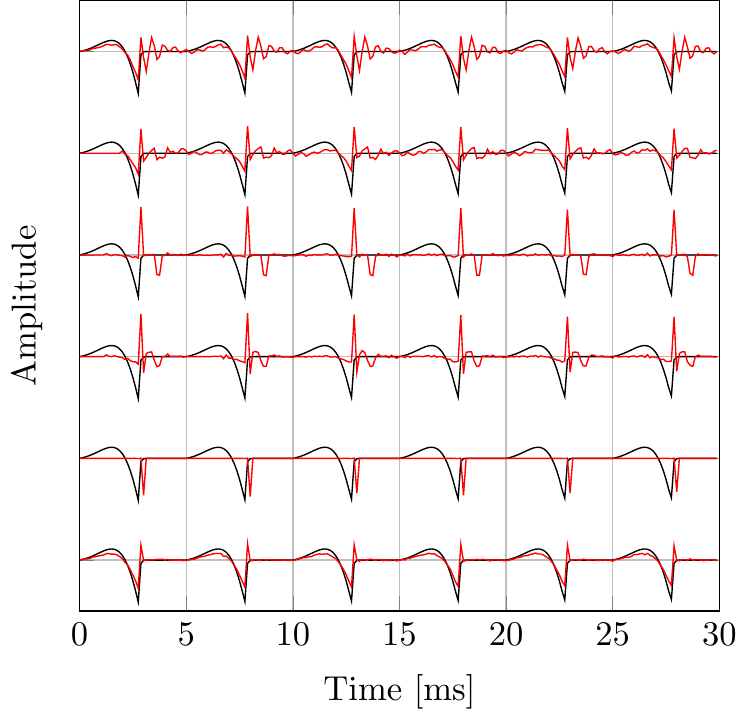}
\caption{residuals estimate for synthetic /n/. The black line is the LF excitation. The red lines shown correspond to, from top to bottom, residuals of (1) 2-norm LP, (2) TS-LS-PZ, (3) 1-norm LP, (4) EM-LP, D=8, (5) VEM-PZ, D=1, (6) VEM-PZ, D=8.}
\label{fig:residual_estimate_synthetic}
\end{figure}

\begin{figure}[tb]
\centering
\includegraphics{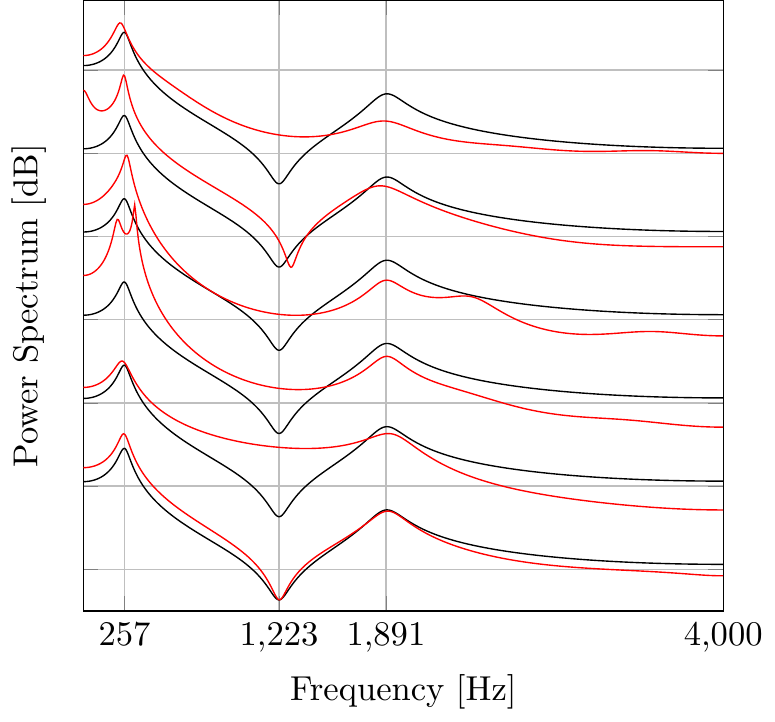}
\caption{corresponding spectral estimates for synthetic /n/. The red lines have the same setting as {Fig.\ 1}.}
\label{fig:spectral_estimate_synthetic2}
\end{figure}
In this section, we compare the performance of the proposed VEM-PZ, the two-stage least squares pole-zero (TS-LS-PZ) method \cite{Stoica2004a}, 2-norm linear prediction (2-norm LP) \cite{Makhoul1975}, 1-norm linear prediction (1-norm LP)\cite{Giacobello2012} and expectation maximization based linear prediction (EM-LP) for mixed excitation \cite{Giri2014} in both synthetic and real speech signals analysis scenarios. 
\subsection{Synthetic signal analysis}
\begin{table}
	\centering
	\caption{The spectral distortion}
	\begin{tabular}{lccccccccc} \toprule[.5pt]\toprule[.5pt]
		\input{figure3_sd_result_test_test.tex}
		\bottomrule[.5pt]\bottomrule[.5pt]
		\label{tab:sd_table1}
	\end{tabular}
\end{table}
\begin{figure}[tb]
\centering
\includegraphics{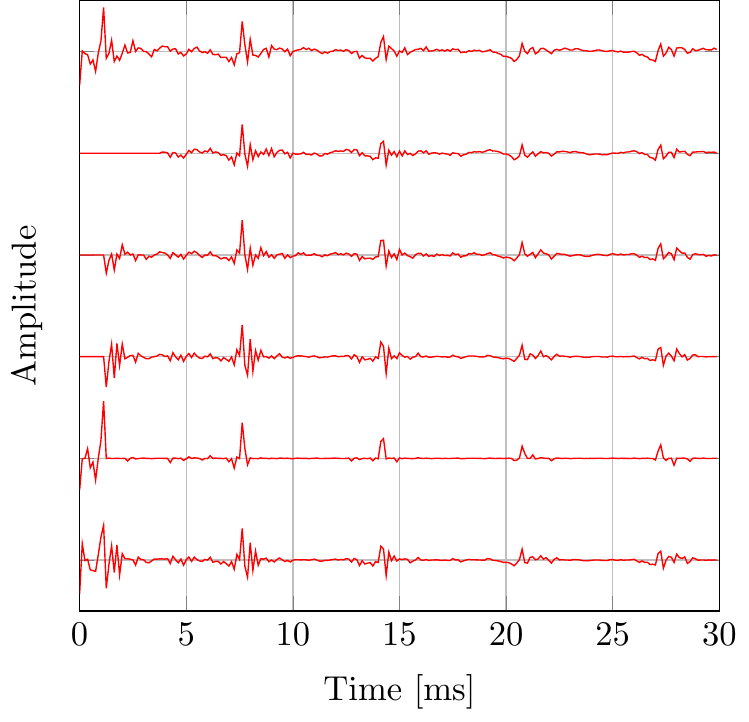}
\caption{residuals estimate for /n/ in the word "manage". The red lines have the same setting as {Fig.\ 1}.}
\label{fig:residual_estimate_real}
\end{figure}

\begin{figure}[tb]
\centering
\includegraphics{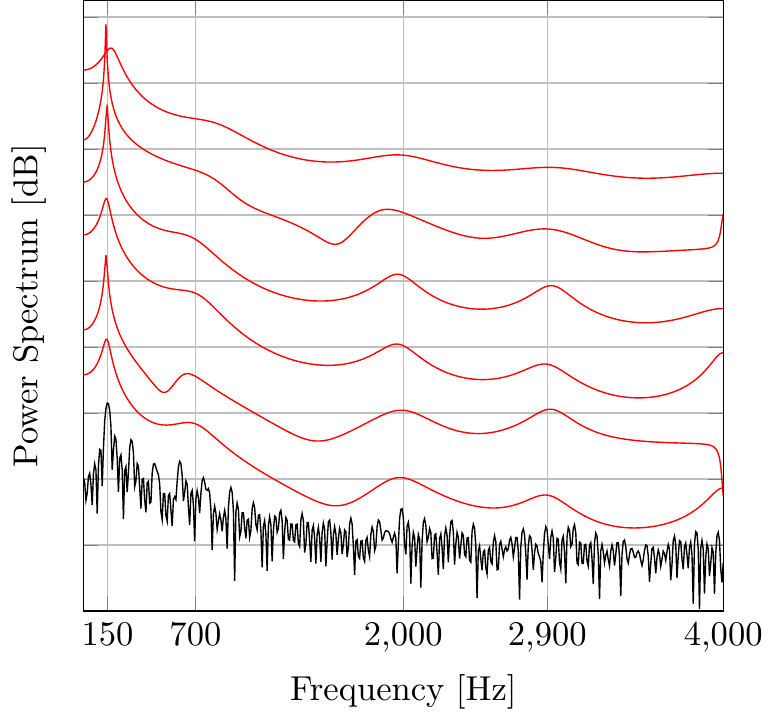}
\caption{corresponding spectral estimates for /n/ in the word "manage". The red lines have the same setting as {Fig.\ 1}. The black line is the periodogram.}
\label{fig:spectral_estimate_real2}
\end{figure}
We first examine the performance of the VEM-PZ with different block size $ {D}$ and compares it with traditional algorithms using synthetic voiced consonant /n/,  as shown in Fig.\ \ref{fig:residual_estimate_synthetic} and Fig.\ \ref{fig:spectral_estimate_synthetic2}.  The synthetic signals are generated by convolving an artificial glottal source waveform with a constructed filter. The first derivative of the glottal flow pulse is simulated with the Liljencrants-Fant (LF) waveform \cite{Fant1985} with the modal phonation mode, whose parameter is taken from Gobl \cite{Gobl1989}. The voiced alveolar nasal consonant /n/ is synthesized at 8 kHz sampling frequency with the constructed filter having two formant frequencies (bandwidths) of 257 Hz (32 Hz) and 1891 Hz (100 Hz) and one antiformant of 1223 Hz (52 Hz) \cite{Mehta2012}. $ {N}$ is set to 240 (30 ms), $e=1$ and $c=d=f=10^{-6}$ are used for all the experiments. The power  ratio of the block sparse excitation $\mathbf{e}$ and Gaussian component $\mathbf{m}$ is set to 30 dB, and the pitch is set to 200 Hz. The $K$ and $L$ are set to 5 for the pole-zero modeling methods (i.e., VEM-PZ and TS-LS-PZ), but $K=10$ is used for the all-pole modeling methods (i.e., 2-norm LP, 1-norm LP and EM-LP). In Fig.\ \ref{fig:residual_estimate_synthetic}, the means of the residuals of EM-LP and VEM-PZ are plotted. Note that the residuals of the TS-LS-PZ, 1-norm LP and EM-LP are prepended with zeros due to the covariance-based estimation methods. As can be seen in Fig.\ \ref{fig:residual_estimate_synthetic}, the residual estimate of the proposed VEM-PZ with $ {D}=8$ is closest to the true block sparse excitation. Moreover, when $ {D}=1$, the residuals of the VEM-PZ are the sparsest as expected. Although the EM-LP also produces block sparse residuals compared with the 1-norm LP, the proposed method achieves the best approximation to the true one due to the usage of the pole-zero modeling and the block-sparse motivated SBL method. The corresponding spectral estimates are shown in Fig.\ \ref{fig:spectral_estimate_synthetic2}. First, as can be seen, the VEM-PZ with $ {D}=1$ has a smooth power spectrum due to the sparse residuals in Fig.\ \ref{fig:residual_estimate_synthetic}. Second, the 1-norm LP tends to produce a better estimate of the formants than the 2-norm LP and TS-LS-PZ. Third, although the EM-LP has two peaks around the first formant, it performs well for second formant estimation. Finally, the proposed VEM-PZ with $ {D}=8$  has good performance for both formant, antiformant and bandwidth estimates because of the block sparse excitation and the pole-zero model.

Second, the spectral distortion is tested under different fundamental frequencies and block sizes. The measure is defined as the truncated power cepstral distance \cite{Rabiner1993}, i.e.,
\begin{align}
 {d_{\mathrm{ceps}}}= \sum_{n=-S}^{S} (c_n-\hat{c}_n)^2,
\end{align}
where $c_n$ and $\hat{c}_n$ are the true and estimated power cepstral coefficients, respectively. Cepstral coefficients are computed by first reflecting all the poles and zeros to the inside of the unit circle and then using the recursive relation in \cite{Mehta2012}. In our experiments, we set $S=300$. The fundamental frequency rises from 200 to 400 Hz in steps of 50 Hz. The experimental results in TABLE \ref{tab:sd_table1} are obtained by the ensemble averages over 500 Monte Carlo experiments.  $ {D}=6$ is used for the EM-LP \cite{Giri2014}. As can be seen from TABLE \ref{tab:sd_table1}, the 2-norm LP has a lower spectral distortion than the 1-norm LP, EM-LP and TS-LS-PZ (except for 300 and 350 Hz). The proposed VEM-PZ achieves the lowest spectral distortion for 200, 300, 350 and 400 Hz. However, note that the good performance of the VEM-PZ depends on a good choice of the block sizes for different fundamental frequencies, and there is a fluctuation when the frequency changes. This is because the length of correlated samples changes with the fundamental frequency. But, as can be seen from Fig.\ \ref{fig:spectral_estimate_synthetic2} and from our experience, the VEM-PZ usually produces better formant, antiformant and bandwidth estimates than traditional ones.
\subsection{Speech signal analysis}
We examine the residuals and spectral estimate of the VEM-PZ for a nasal consonant /n/ in the word "manage" from the CMU Arctic database \cite{database, databaseurl}, pronounced by an US female native speaker, sampled of 8000 Hz. The results are shown in Fig.\ \ref{fig:residual_estimate_real} and Fig.\ \ref{fig:spectral_estimate_real2}. To improve the modeling flexibility, the $K$ and $L$ are set to 10 for the PZ methods (i.e., VEM-PZ and TS-LS-PZ), but $K=10$ is still used for the all-pole methods (i.e., 2-norm LP, 1-norm LP and EM-LP).  As can be seen from Fig.\ \ref{fig:residual_estimate_real}, the residuals of the EM-LP and 1-norm LP are sparser than the 2-norm LP. The residuals of the proposed VEM-PZ with $D=1$ are the sparsest. The proposed VEM-PZ with $D=8$ is block sparse and is sparser than the TS-LS-PZ. From Fig.\ \ref{fig:spectral_estimate_real2}, we can see that all the algorithms have formant estimates around 150 Hz. However, the TS-LS-PZ, 1-norm LP and VEM-PZ with $D=1$ have very peaky behaviour. Also, the 2-norm LP produces bad bandwidth estimates around 2000 and 2900 Hz. Furthermore, compared to the EM-LP, the proposed VEM-PZ with D=8 has good antiformant estimates around 500 and 1500 Hz. 
\section{Conclusion}
A variational expectation maximization pole-zero speech analysis method has been proposed. By using the pole-zero model, it can fit the spectral zeros of speech signals easily. Moreover, block sparse residuals are encouraged by applying the sparse Bayesian learning method. By iteratively updating parameters and statistics of residuals and hyperparameters, block sparse residuals can be obtained. The good performance has been verified by both synthetic and real speech experiments. The proposed method is promising for speech analysis applications. Next, further research into the formant, antiformant and bandwidth estimation accuracy, stability, and unknown sparse pattern should be conducted.



%
\bibliographystyle{IEEEbib_style}
\bibliography{IEEEabrv,myabrv,eusipco}

\end{document}

%% file: figure3_sd_result_test_test.tex
F0	 &200	 &250	 &300	 &350	 &400	  \\	\midrule[.5pt]
2-norm LP	 	 &1.79	 &\textbf{2.14}	 &2.12	 &2.53	 &2.13	  \\
TS-LS-PZ		 &2.41	 &4.77	 &1.88	 &1.46	 &2.86	 	 \\
1-norm LP	 &2.43	 &3.15	 &3.60	 &3.29	 &4.29	 	 \\
EM-LP	  &5.62	 &6.68	 &4.68	 &3.96	 &4.83	 \\
VEM-PZ, D=1		 &4.50	 &7.14	 &2.29	 &1.54	 &2.31	 	 \\
VEM-PZ, D=5 &1.55	 &4.47	 &\textbf{0.69}	 &2.01	 &4.50		 \\
VEM-PZ, D=7	 	 &2.08	 &4.07	 &2.18	 &\textbf{1.41}	 &1.29	 	 \\
VEM-PZ, D=8		 &\textbf{0.77}	 &5.56	 &2.52	 &4.86	 &\textbf{0.53}	 	\\ 